\documentclass[12pt]{article}

\def\ov{\overline}

\def\pa{\partial}
\def\la{\langle}
\def\ra{\rangle}
\def\be{\begin{equation}}
\def\ee{\end{equation}}

\newtheorem{theorem}{Theorem}
\newtheorem{lemma}[theorem]{Lemma}

\newtheorem{definition}[theorem]{Definition}

\newtheorem{proposition}[theorem]{Proposition}

\begin{document}

\title{Multidimensional ultrametric \\ pseudodifferential equations}

\author{S.Albeverio, S.V.Kozyrev}

\maketitle

\begin{abstract}
We develop an analysis of wavelets and pseudodifferential operators
on multidimensional ultrametric spaces which are defined as products
of locally compact ultrametric spaces. We introduce bases of
wavelets, spaces of generalized functions and Lizorkin generalized
functions on multidimensional ultrametric spaces.

We also consider some family of pseudodifferential operators on
multidimensional ultrametric spaces. The notions of Cauchy problem
for ultrametric pseudodifferential equations and of ultrametric
characteristics are introduced. A theorem about existence and
uniqueness of the solution for the Cauchy problem (the analogue of
the Kovalevskaya theorem) is proven.
\end{abstract}


\bigskip

AMS--classification: 26E30, 26A33, 45N05, 45E99, 42C40, 46F12

\section{Introduction}

Ultrametric and $p$--adic analysis in interaction with mathematical
physics constitutes a broad field of research, see e.g. \cite{VVZ},
\cite{Andr}, \cite{Kochubei}, \cite{RandomWalk}, \cite{AlbZhao},
\cite{Yasuda}, \cite{Evans}. We mention the $p$--adic wavelets
introduced in \cite{wavelets} and generalized in \cite{Benedetto} to
some family of abelian groups. The generalization of $p$--adic
wavelets to the case of underlying general locally compact
ultrametric spaces  \cite{Izv}, \cite{ACHA}, \cite{MathSbornik} is a
crucial tool for the approach of the present paper. In \cite{ShSk}
new examples of $p$--adic wavelets were constructed and it was shown
that the basis of $p$--adic Haar wavelets is not unique. In
\cite{contwavelets} it was shown that the continuous $p$--adic
wavelet transform coincides with the expansion in the series on
$p$--adic wavelets from the basis constructed in \cite{wavelets}.

An important notion of ultrametric analysis is one of the
ultrametric pseudodifferential operators, defined by the following
formula, see \cite{Izv}, \cite{ACHA}, \cite{MathSbornik},
\cite{nhoper}, (and also the Appendix 2, for details)
$$
Tf(x)=\int_{X}T({\rm sup}(x-y))(f(x)-f(y))d\nu(y).
$$
The above operator acts in the space $L^2(X,\nu)$ of complex valued
functions which are quadratically integrable with respect to the
measure $\nu$ on a locally compact ultrametric space $X$. It was
found \cite{Izv}, \cite{ACHA}, \cite{MathSbornik} that operators of
this form are diagonal with respect to a natural basis consisting of
ultrametric wavelets in $L^2(X,\nu)$, and one can compute the
spectra of these operators using an explicit formula. Moreover, the
eigenvalues of the above operator are in one--to--one correspondence
with the balls of non--zero diameters in the space $X$.

In the present paper we investigate several topics of ultrametric
analysis.

First, we consider multidimensional ultrametric spaces, which are
defined as products $X=X^1\times \dots \times X^n$ of locally
compact ultrametric spaces with the product measure $\nu=\nu^1\times
\dots \times\nu^n$. We introduce bases of wavelets, spaces of
generalized functions and Lizorkin's generalized functions on
multidimensional ultrametric spaces.

The space of Lizorkin generalized functions in the $p$--adic case
was introduced in \cite{AKhSh}. We generalize this definition to the
general ultrametric case. In particular, we find the
characterization of Lizorkin generalized functions by formal series
of ultrametric wavelets.

Second, we consider the family of pseudodifferential operators on
multidimensional ultrametric spaces, which contains polynomials in
one--dimensional ultrametric pseudodifferential operators. These
operators are diagonal in the mentioned above bases of
multidimensional ultrametric wavelets, and moreover, the eigenvalues
are in one--to--one correspondence with the products of balls ${\bf
I}=I^1\times\dots\times I^n$, where $I^j$ is a ball in $X^j$. We
introduce the notion of ultrametric characteristics --- we say that
the product of balls ${\bf I}$ is characteristic for the operator
$T$ if the corresponding eigenvalue $\lambda_{\bf I}$ of the
operator $T$ is equal to zero.

Third, we introduce the following notion of Cauchy problem for
ultrametric pseudodifferential equations: we say that $u$ is a
solution of Cauchy problem
$$
Tu=f
$$
if $T$ is an ultrametric pseudodifferential operator, $f$ lies in
Lizorkin space of generalized functions, $u$ lies in the space of
generalized functions and satisfies the initial condition
$u(\chi_{\bf I})=u_0\nu({\bf I})$, where $\chi_{\bf I}$ is the
characteristic function of ${\bf I}$ and $u_0$ is a complex number.

We prove theorem about existence of the solution for the Cauchy
problem (the analogue of the Kovalevskaya theorem). We show that the
Cauchy problem possesses a unique solution if there are no
characteristics.

The main technical tool for the investigations of the present paper
is the expansion of generalized function on an ultrametric space to
a series of ultrametric wavelets, given by Lemma \ref{lemma_series}
in the one dimensional case and by Lemma \ref{lemma_series_1} in the
multidimensional case.

The structure of the present paper is as follows.

In section 2 we discuss Lizorkin space of generalized functions on
an ultrametric space and show that these generalized functions can
be considered as formal series over ultrametric wavelets.

In section 3 we find expansion of a generalized function on a
regular ultrametric space in a series over wavelets, introduce the
notion of Cauchy problem for pseudodifferential equations on an
ultrametric space, and prove the theorem about the existence and
uniqueness of a solution for the Cauchy problem.

In section 4 we build bases of wavelets on multidimensional
ultrametric spaces and discuss the corresponding hypergraphs.

In section 5 we discuss generalized functions on multidimensional
ultrametric spaces.

In section 6 we find expansion of generalized functions in series of
ultrametric wavelets in the multidimensional case.

In section 7 we introduce a family of multidimensional ultrametric
pseudodifferential operators and introduce the notion of an
ultrametric characteristic.

In section 8 we prove the theorem about existence and uniqueness of
the solution for Cauchy problem for multidimensional ultrametric
pseudodifferential equations (the analogue of the Kovalevskaya
theorem).

In section 9 (Appendix 1) we recall the notion of a characteristic
for  (real) equations in partial derivatives.

In section 10 (Appendix 2) we recall some results on analysis of
wavelets and pseudodifferential operators on ultrametric spaces.

\section{Lizorkin space of generalized functions}

Consider the space $D_0(X)\subset D(X)$ of mean zero test functions
on a regular ultrametric space $X$. This space possesses the
filtration by subspaces $D_0({\cal S})$ of mean zero test functions
in $D({\cal S})$. It is easy to see that the space $D_0({\cal S})$
is the linear span of wavelets $\Psi_{Ij}$ where $I\in {\cal S}$
runs over all non--minimal vertices in ${\cal S}$. The space
$D_0(X)$ is spanned by all ultrametric wavelets. The topology on
$D_0(X)$ is introduced in the obvious way.

\begin{definition}\label{Lizorkin}{\sl
Lizorkin space $D'_0(X)$ of generalized functions is the space of
linear functionals on the space $D_0(X)$ of mean zero test
functions.}
\end{definition}

We have the following two characterizations of Lizorkin space.

1) Since the space $D_0(X)$ is exactly the space of test functions
which lie in the kernel of the constants (we consider the constants
as generalized functions), the space $D'_0(X)$ coincides with the
space of equivalence classes of generalized functions which are
equal up to addition of a constant. In particular, any generalized
function in $D'(X)$ can be considered also as a generalized function
in Lizorkin space $D'_0(X)$.

2) Since the space $D_0(X)$ is a space of finite linear combinations
of ultrametric wavelets from the orthonormal basis $\{\Psi_{Ij}\}$
in $L^2(X,\nu)$, the space $D'_0(X)$ of linear functionals on
$D_0(X)$ can be identified with the space of formal series over
ultrametric wavelets, with the following action on $D_0(X)$.

Let $f\in D_0(X)$ have the form of a finite linear combination of
wavelets
$$
f=\sum_{Ij}f_{Ij}\Psi_{Ij},\qquad f_{Ij}=\la \Psi_{Ij}, f \ra,
$$
where $\la\cdot,\cdot \ra$ is the scalar product in $L^2(X,\nu)$:
$$
\la f ,g \ra=\int_{X}\ov{f(x)}g(x)d\nu(x)
$$
where $\ov{f}$ denotes the complex conjugate to $f$.

Let $\phi\in D'_0(X)$ have the form of a series over wavelets:
$$
\phi=\sum_{I'j'}\phi_{I'j'}\Psi_{I'j'},\qquad
\phi_{I'j'}=\phi(\ov{\Psi_{I'j'}}).
$$

Then the action of $\phi$ on $f$ is given by the following finite
linear combination
$$
\phi(f)=\sum_{Ij}\phi_{Ij}f_{Ij}.
$$

\section{Cauchy problem in the one dimensional case}

In the previous section we discussed expansion of generalized
function from Lizorkin space $D'_0(X)$ in series over ultrametric
wavelets. In the present section we build the analogous expansion
for generalized functions in $D'(X)$.

\begin{lemma}\label{fix_gf}
{\sl Assume that for a ball $I_0\subset X$ its $\nu$--measure is
positive: $\nu(I_0)>0$.

Then there exists the unique generalized function $u\in D'(X)$ such
that
$$
u\left(\chi_{I_0}\right)=u_0\nu(I_0),\qquad
u\left(\ov{\Psi_{Ij}}\right)=u_{Ij},\quad\forall I,j.
$$
 }
\end{lemma}

\noindent{\it Proof}\qquad We note that for any regular subtree
${\cal S}\subset {\cal T}(X)$ (see the Appendix 2 for the notations)
the corresponding space $D(\cal S)$ of test functions is an
orthogonal sum of the space $D_0(\cal S)$ of mean zero test
functions (spanned by the corresponding wavelets) and the one
dimensional linear space spanned by the characteristic function
$\chi_{{\rm sup}({\cal S})}$ of the ball ${\rm sup}({\cal S})$. The
ball ${\rm sup}({\cal S})$ is the largest ball in the set ${\cal S}$
and is the support of $D(\cal S)$.

If a ball $I_0\in {\cal S}$ has positive measure $\nu(I_0)>0$, then
its characteristic function $\chi_{I_0}$ is not orthogonal to the
characteristic function $\chi_{{\rm sup}({\cal S})}$. Therefore the
union of the basis of wavelets in $D_0(\cal S)$ and the
characteristic function $\chi_{I_0}$ is a complete set in $D(\cal
S)$.

Therefore the action of an arbitrary generalized function $u$ in
$D(\cal S)$ can be defined by its application to all wavelets and to
a characteristic function $\chi_{I_0}$, $\nu(I_0)>0$. This finishes
the proof of the lemma.

\bigskip

Let us prove the following important lemma about the expansion of
generalized functions in $D'(X)$ into series over wavelets.

\begin{lemma}\label{lemma_series}
{\sl Assume that for a ball $I_0\subset X$  its $\nu$--measure is
positive: $\nu(I_0)>0$.

Then the series \be\label{the_series}
u=u_0+\sum_{Ij}u_{Ij}\left(\Psi_{Ij}-{1\over\nu(I_0)}\Psi_{Ij}\left(\chi_{I_0}\right)\right)
\ee is the generalized function $u\in D'(X)$ which satisfies the
condition \be\label{init_cond}
u\left(\chi_{I_0}\right)=u_0\nu(I_0),\qquad
u\left(\ov{\Psi_{Ij}}\right)=u_{Ij}. \ee

Moreover an arbitrary $u\in D'(X)$ satisfying (\ref{init_cond}) has
the form (\ref{the_series}). }
\end{lemma}

\noindent{\it Proof}\qquad To check that the series
(\ref{the_series}) is indeed a generalized function in $D'(X)$, we
have to prove that application of this series to any characteristic
function $\chi_{J_0}$ gives a convergent series. We have
$$
u\left(\chi_{J_0}\right)=u_0\nu(J_0)+\sum_{Ij}u_{Ij}
\left(\Psi_{Ij}\left(\chi_{J_0}\right)-{\nu(J_0)\over\nu(I_0)}\Psi_{Ij}\left(\chi_{I_0}\right)\right)
$$
We see that for sufficiently large $I$, namely for $I>{\rm
sup}\,(J_0,I_0)$, we get
$$
\nu^{-1}(I_0)\Psi_{Ij}\left(\chi_{I_0}\right)=\Psi_{Ij}(x)
$$
where $x\in I_0$. Analogously
$$
\Psi_{Ij}\left(\chi_{J_0}\right)=\nu(J_0)\Psi_{Ij}(x')
$$
where $x'\in J_0$.

Since for $I>{\rm sup}\,(J_0,I_0)$ we have
$\Psi_{Ij}(x)=\Psi_{Ij}(x')$, the corresponding terms in the series
above cancel, and the series will take the form of the finite sum
$$
u\left(\chi_{J_0}\right)=u_0\nu(J_0)+\sum_{Ij:J_0<I\le{\rm
sup}\,(J_0,I_0)}u_{Ij}
\Psi_{Ij}\left(\chi_{J_0}\right)-{\nu(J_0)\over\nu(I_0)}\sum_{Ij:I_0<I\le{\rm
sup}\,(J_0,I_0)}u_{Ij}\Psi_{Ij}\left(\chi_{I_0}\right).
$$
This proves that the series (\ref{the_series}) is a generalized
function in $D'(X)$.

The application of (\ref{the_series}) on $\chi_{I_0}$ proves that
$u$ satisfies the condition $u\left(\chi_{I_0}\right)=u_0\nu(I_0)$.

Analogously, the application of (\ref{the_series}) to
$\ov{\Psi_{Ij}}$, taking into account that
$$
\Psi_{I'j'}\left(\ov{\Psi_{Ij}}\right)=\la\Psi_{Ij},\Psi_{I'j'}\ra=\delta_{II'}\delta_{jj'},
$$
proves that $u$ satisfies the condition
$u_{Ij}=u\left(\ov{\Psi_{Ij}}\right)$.

Then by applying Lemma \ref{fix_gf} we prove that an arbitrary $u\in
D'(X)$ satisfying (\ref{init_cond}) has the form (\ref{the_series}).
This finishes the proof of the lemma.

\bigskip

\noindent{\bf Remark}\qquad We understand the series
(\ref{the_series}) as a limit with respect to a filtration, i.e. a
finite sum of the series depends on the regular subtree ${\cal S}$,
and the series is a limit of finite sums with ${\cal S}\to{\cal
T}(X)$.

\bigskip

Let us consider the ultrametric pseudodifferential operator $T$ on
the regular ultrametric space $X$ with the eigenvalues $\lambda_{
I}$ in the wavelet basis.

For an arbitrary generalized function $u\in D'(X)$ the application
of a pseudodifferential operator $T$ to this function gives a
generalized function in Lizorkin space $Tu\in D'_0(X)$, with the
action on wavelets defined as follows
$$
Tu(\ov{\Psi_{Ij}})=u(\ov{T^*\Psi_{Ij}})=\lambda_I u(\ov{\Psi_{Ij}}).
$$
This formula is a direct analogue of the well known formula of real
analysis which defines the derivative of a generalized function by
$u'(\phi)=-u(\phi')$, where $u$ is a generalized function and $\phi$
is a test function on the real line.

Therefore ultrametric pseudodifferential operators can be considered
as linear maps $D'(X)\to D'_0(X)$, or, in the analogous way, as
linear maps $D'_0(X)\to D'_0(X)$. Note that in general it is not
possible to consider a pseudodifferential operator as a map
$D'(X)\to D'(X)$.

\begin{definition}{\sl We say that a generalized function $u\in D'(X)$ is
a solution of Cauchy problem if \be\label{Cauchy1} Tu=f, \ee where
$f\in D'_0(X)$ lies in Lizorkin space, and $u$ satisfies the initial
condition: $u\left(\chi_{I_0}\right)=u_0\nu(I_0)$, for the
characteristic function $\chi_{I_0}$ of some ball $I_0\subset X$,
$\nu(I_0)>0$. }
\end{definition}

It is easy to see that there exist necessary conditions for
solvability of (\ref{Cauchy1}): if $\lambda_I=0$ for some $I$ then
all $f_{Ij}=f\left(\ov{\Psi_{Ij}}\right)=0$.

\begin{theorem}
{\sl Let the Lizorkin generalized function $f$ at the RHS of
(\ref{Cauchy1}) satisfy the necessary conditions
$f_{Ij}=f\left(\ov{\Psi_{Ij}}\right)=0$ for $I$ for which
$\lambda_I=0$ and for all corresponding indices $j$.

Then there exists a solution of the Cauchy problem (\ref{Cauchy1}),
and an arbitrary solution $u$ of the Cauchy problem (\ref{Cauchy1})
will be given by the formula \be\label{solution}
u=u_0+\sum_{Ij}{1\over\lambda_I}f_{Ij}\left(\Psi_{Ij}-{1\over\nu(I_0)}\Psi_{Ij}\left(\chi_{I_0}\right)\right)
+\sum_{Jj}u_{Jj}\left(\Psi_{Jj}-{1\over\nu(I_0)}\Psi_{Jj}\left(\chi_{I_0}\right)\right).
\ee The summation runs over $I$ for which $\lambda_I\ne 0$ and over
$J$ for which $\lambda_J=0$. Here
$f_{Ij}=f\left(\ov{\Psi_{Ij}}\right)$ and $u_{Jj}$ are arbitrary.}
\end{theorem}

\noindent{\it Proof}\qquad By Lemma \ref{lemma_series} a generalized
function $u\in D'(X)$ satisfying the initial condition
$u\left(\chi_{I_0}\right)=u_0\nu(I_0)$ has the form
(\ref{the_series}).

Since the operator $T$ is diagonal in the basis of wavelets,
applying $T$ to (\ref{the_series}) we get from (\ref{Cauchy1}) for
$I$, for which $\lambda_I\ne 0$
$$
u_{Ij}={1\over\lambda_I}f_{Ij}.
$$

Application of the operator $T$ to (\ref{solution}) proves that $u$
is a solution of (\ref{Cauchy1}). This finishes the proof of the
theorem.

\section{Product of ultrametric spaces and wavelets}

Assume we have several regular ultrametric spaces $X^1,\dots,X^n$
with Borel measures $\nu^1,\dots,\nu^n$. Let us consider the product
space $X=X^1\times \dots \times X^n$ with the product measure
$\nu=\nu^1\times \dots \times\nu^n$.

Each of the spaces $X^i$ is dual to the corresponding directed tree
${\cal T}(X^i)$. The tree ${\cal T}(X^i)$ is defined by the set of
vertices (balls), and a family of two--element subsets (edges). The
tree ${\cal T}(X^i)$ is used in the construction of the
corresponding basis of ultrametric wavelets.

In the present section we construct an object, called a hypergraph,
which we will use in the procedure of construction of the basis of
multidimensional wavelets on the space $X=X^1\times \dots \times
X^n$. In general, a hypergraph is a set with the elements called
vertices, and several families of subsets: family of pairs, family
of $k$--tuples, etc. The family of subsets containing $k$ elements
called the family of edges of the order $k-1$. In particular, a
graph is a hypergraph with the family of pairs of vertices, called
the family of edges (of the order one).

We define the hypergraph ${\cal T}={\cal T}(X^1)\times \dots \times
{\cal T}(X^n)$ as follows. We have the set of vertices, which is the
set of $n$--tuples ${\bf I}=(I^1,\dots,I^n)$, where $I^j$ is a
vertex in ${\cal T}(X^j)$. We will also use the notation ${\bf
I}=I^1\times\dots\times I^n$. We denote by $\chi_{\bf I}$ the
function on $X=X^1\times\dots\times X^n$ which is equal to the
characteristic function of the product of balls ${\bf I}$.

Then, we have the set of one--dimensional edges --- the set of pairs
$({\bf I},{\bf J})$, where
$$
{\bf I}=(I^1,\dots, I^k,\dots I^n), \qquad {\bf J}=(I^1,\dots,
J^k,\dots I^n)
$$
for some $k$ (i.e. one element in the tuples differs), and, moreover
$I^k$ and $J^k$ belong to some edge in ${\cal T}(X^k)$. One can say
that the edge $({\bf I},{\bf J})$ can be identified with the direct
product of the edge $(I^k,J^k)$ in ${\cal T}(X^k)$ and the product
of vertices $I^1\times\dots I^{k-1}\times I^{k+1}\dots I^n$ in
${\cal T}(X^j)$, $j\ne k$. Since all edges ${\cal T}(X^k)$ are
ordered sets, the same will hold for the edge $({\bf I},{\bf J})$ in
${\cal T}$. This defines our partial order in the hypergraph ${\cal
T}$.

The set of two--dimensional edges contains quadruples $({\bf A},{\bf
B},{\bf C},{\bf D})$, where $({\bf A},{\bf B})$, $({\bf A},{\bf
C})$, $({\bf B},{\bf D})$ and $({\bf C},{\bf D})$ are
one--dimensional edges. Thus a two--dimensional edge is a square
where the vertices are vertices in ${\cal T}$ and the sides are
one--dimensional edges in ${\cal T}$. In more details this
construction is as follows: let
$$
{\bf A}=(A^1,\dots, A^i,\dots,A^j,\dots, A^n),
$$
$$
{\bf B}=(A^1,\dots, B^i,\dots, A^j,\dots, A^n),
$$
(since $({\bf A},{\bf B})$ is a one--dimensional
edge) and
$${\bf C}=(A^1,\dots,A^i,\dots, C^j,\dots, A^n),$$
then
$${\bf D}=(A^1,\dots, B^i,\dots, C^j,\dots, A^n).$$

Analogously, a $d$--dimensional edge in ${\cal T}$, $d=0,1,\dots,n$
is a $d$--dimensional cube with the sides being $d-1$--dimensional
edges in ${\cal T}$. It is easy to see that a $d$--dimensional edge
in ${\cal T}$ has the order $2^{d}-1$ (i.e. contains $2^d$
vertices).

By construction, since a $d$--dimensional edge is a product of the
one dimensional edges which are ordered sets, it possesses a natural
partial order. A $d$--dimensional edge contains the largest and the
smallest vertices. These vertices are the opposite vertices in the
cube (i.e. the line connecting these vertices is the main diagonal
in the cube). The different edges may intersect and the partial
orders in the intersection in both edges coincide.

We introduce the following notion: we say that the vertex ${\bf
I}=(I^1,\dots, I^n)$ is sufficiently larger than the vertex ${\bf
J}=(J^1,\dots, J^n)$ and we write:
$$
{\bf I}>>{\bf J}
$$
if $\forall i=1,\dots,n$ we have
$$
I^i>J^i.
$$

For any two vertices ${\bf I}=(I^1,\dots, I^n)$ and ${\bf
J}=(J^1,\dots, J^n)$ there exists the unique supremum, i.e. the
minimal vertex that is larger than both ${\bf I}$ and ${\bf J}$:
$$
{\rm sup}({\bf I},{\bf J})=({\rm sup}(I^1,J^1),\dots, {\rm
sup}(I^n,J^n))
$$
Therefore ${\cal T}$ is a directed set.

For any two vertices in ${\cal T}$ there exists a path (i.e. a
sequence of one--dimensional edges) in ${\cal T}$, connecting the
edges. This path in general is not unique.

Let us remind that the branching index $p_I$ of a vertex $I$ in the
tree of balls ${\cal T}(X)$, corresponding to a regular ultrametric
space $X$, is equal to the number of maximal subballs in the ball
$I$ in $X$.

Analogously, we will say that the branching index of the vertex
${\bf I}=(I^1,\dots, I^k,\dots I^n)$ in ${\cal T}$ is equal to
$(p_{I^1},\dots,p_{I^n})$, where $p_{I^j}$ is the branching index of
$I^j$.

The $k$--dimensional edge with the beginning in ${\bf I}$ is called
decreasing if ${\bf I}$ is the maximal vertex in the edge. The
number of decreasing edges of maximal dimension with the beginning
in the vertex ${\bf I}=(I^1,\dots, I^k,\dots I^n)$ is equal to the
product of branching indices
$$
\prod_{j}p_{I^j},
$$
the product being taken over all non--minimal $I^j$ in ${\bf I}$
(with non--zero branching indices). The dimension of decreasing
edges with maximal dimension which begin in ${\bf I}$ is equal to
the number of non--minimal $I^j$ in ${\bf I}$.

If all $I^j$ in ${\bf I}$ are non--minimal, we will say that the
vertex $I$ is generic. In this case decreasing edges of maximal
dimension with the beginning in ${\bf I}$ have the dimension $n$.

Let us discuss the wavelet space corresponding to a vertex ${\bf I}$
in the hypergraph ${\cal T}$. If ${\bf I}=(I^1,\dots, I^k,\dots
I^n)$ is a generic vertex, then we define the wavelet space
$V^0({\bf I})$ as the tensor product of wavelet spaces corresponding
to vertices $I^j$ in ${\cal T}(X^j)$:
$$
V^{0}({\bf I})=\otimes_{j} V^{0}(I^j).
$$

The space $V^0({\bf I})$ possesses the orthonormal basis consisting
of tensor products of wavelets \be\label{multiwavelet} \Psi_{{\bf I}
{\bf j}}=\otimes_{k}\Psi_{I^k j^k} \ee where $\{\Psi_{I^k j^k}\}$ is
the orthonormal basis in the wavelet space corresponding to the
vertex $I^k\in {\cal T}(X^k)$, ${\bf I}=(I^1,\dots,I^n)$, ${\bf
j}=(j^1,\dots,j^n)$.

We will use the hypergraph ${\cal T}(X)$ to construct the wavelet
bases in $L^2(X,\nu)$. We will get the following problem. It might
happen that some of the spaces $X^i$ could have a finite volume
$\nu^i(X^i)<\infty$. In this case, in order to obtain basis in
$L^2(X^i,\nu^i)$, we should add a constant function to the set of
wavelets, corresponding to non-minimal balls $I^i\in {\cal T}(X^i)$.
The wavelets $\Psi_{Ij}$ correspond to the ball $I\in {\cal
T}(X^i)$, but the constant function in general can not be put into
correspondence to any of the balls in ${\cal T}(X^i)$.

Let us perform the following construction. If the volume
$\nu^i(X^i)=A>0$ is finite, let us add to the tree ${\cal T}(X^i)$
of non--minimal balls the vertex $K^{(i)}$, which is larger than all
vertices in ${\cal T}(X^i)$. We denote by $\widetilde {\cal
T}(X^i)={\cal T}(X^i)\bigcup K^{(i)}$, this is a directed set (but
as a graph this set is not connected, because we do not assume
existence of edges containing vertex $K^{(i)}$). Then the basis of
ultrametric wavelets in $L^2(X^i,\nu^i)$ could be described as a
union of wavelets corresponding to vertices in ${\cal T}(X^i)$, and
a constant function $\Psi_{K^{(i)}}= A^{-1/2}$ in $L^2(X^i,\nu^i)$
which corresponds to vertex $K^{(i)}$ (we consider this vertex as
generic). If the measure of $X^i$ is infinite, we define $\widetilde
{\cal T}(X^i)={\cal T}(X^i)$.

We will call the augmented hypergraph $\widetilde {\cal T}$ the
hypergraph
$$
\widetilde {\cal T}=\widetilde {\cal T}(X^1)\times\dots\times
\widetilde {\cal T}(X^n).
$$
The edges and partial order in the augmented hypergraph are
introduced in the same way as in the hypergraph ${\cal T}$. Let us
note that the augmented hypergraph $\widetilde {\cal T}$, in
general, is not connected.

We consider multidimensional wavelets defined by the formula
(\ref{multiwavelet}), whose vertices ${\bf I}$ belong to the
augmented hypergraph $\widetilde {\cal T}$. Let us formulate the
following theorem about the multidimensional basis of ultrametric
wavelets.

\begin{theorem}
{\sl 1) Let the volumes $\nu^i(X^i)$, $i=1,\dots,n$ be infinite.
Then the union $\{\Psi_{{\bf I} {\bf j}}\}$ of the bases in
$V^0({\bf I})$ over all generic ${\bf I}\in {\cal T}$ is an
orthonormal basis in $L^2(X,\nu)$.

2) Assume that some of the volumes $\nu^i(X^i)=A_i$ are finite. Then
the set $\{\Psi_{{\bf I} {\bf j}}\}$, where ${\bf I}$ runs over
generic vertices in the augmented hypergraph $\widetilde {\cal T}$
is an orthonormal basis in $L^2(X,\nu)$. }
\end{theorem}

\section{Multidimensional generalized functions}

For the multidimensional ultrametric space $X=X^1\times\dots\times
X^n$ we define the space of test functions $D(X)$ as the tensor
product of one--dimensional spaces of test functions:
$$
D(X)=\otimes_{i=1}^n D(X^i)
$$
Analogously, we define Lizorkin space of test functions $D_0(X)$ as
$$
D_0(X)=\otimes_{i=1}^n D_0(X^i)
$$
Equivalently, the space $D_0(X)$ can be defined as the space of test
functions satisfying
$$
\int_{X^i} f(x^1,\dots,x^i,\dots,x^n) d\nu^i(x^i)=0,\qquad \forall
i=1,\dots,n.
$$

Topologies in the spaces $D(X)$ and $D_0(X)$ are defined in the
standard way. Since $D(X^i)$ are filtrated by $D({\cal S}^i)$, the
space $D(X)$ will be filtrated by finite dimensional subspaces
$$
D({\cal
S}^1\times\dots\times {\cal S}^n)=D({\cal S}^1)\otimes\dots\otimes
D({\cal S}^n).
$$
A sequence in $D(X)$ converges if it belongs to some subspace
$D({\cal S})$ and converges as a sequence in this finite dimensional
space. The analogous statements will hold for $D_0(X)$.

The space $D'(X)$ of generalized functions is the space of linear
functionals on $D(X)$. The space $D'(X)$ possesses the topology of
weak convergence of the functionals.

Lizorkin space of generalized functions $D'_0(X)$ is the space of
linear functionals on $D_0(X)$ with the topology of weak
convergence.

Since the space $D_0(X)$ can be identified with the space of finite
linear combinations of multidimensional wavelets $\Psi_{\bf Ij}$,
${\bf I}\in {\cal T}$, Lizorkin space $D'_0(X)$ can be considered as
the space of series over wavelets
$$
f=\sum_{{\bf Ij}, {\bf I}\in {\cal T}}f_{\bf Ij}\Psi_{\bf Ij},\qquad
f_{\bf Ij}=f\left(\ov{\Psi_{\bf Ij}}\right).
$$
Here ${\bf I}$ runs over generic vertices in the hypergraph ${\cal
T}$.

Analogously, Lizorkin space of generalized functions can be
identified with the factorspace $D'(X)/D_0^{\bot}(X)$, where
$D_0^{\bot}(X)$ is the subspace of generalized functions which kill
all test functions in $D_0(X)$.

\section{Expansion of generalized functions over wavelets}

\begin{lemma}\label{fix_gf_1}
{\sl Assume that for a vertex ${\bf I}_0\in {\cal T}$, ${\bf
I}_0=I_0^1\times\dots\times I_0^n$ its measure is positive:
$\nu({\bf I}_0)=\prod_{i=1}^{n}\nu^i(I_0^i)>0$.

Denote by $\Psi_{I^ij^i}$, $I^i\in {\cal T}(X^i)$,
$j^i=0,1,\dots,p_{I^i}-1$ the function on the ultrametric space
$X^i$, which is defined as follows:

1) it is equal to the corresponding ultrametric wavelet for
$j^i=1,\dots,p_{I^i}-1$ and non--minimal $I^i\in{\cal T}(X^i)$;

2) it is equal to zero for $j^i=1,\dots,p_{I^i}-1$ and minimal
$I^i\in{\cal T}(X^i)$;

3) it is equal to $\chi_{I_0^i}$ for $j^i=0$ and $I^i=I_0^i$;

4) it is equal to zero for $j^i=0$ and $I^i\ne I_0^i$).

Define for ${\bf I}\in {\cal T}$ and ${\bf j}=(j^1,\dots,j^n)$,
$j^i=0,1,\dots,p_{I^i}-1$
$$
\Psi_{\bf Ij}=\otimes_{i=1}^n\Psi_{I^ij^i}
$$

Then:

1) the set $\{\Psi_{\bf Ij}\}$ is a linearly independent and is a
complete set in $D(X)$;

2) there exists a unique generalized function $u\in D'(X)$ taking
the values
$$
u\left(\ov{\Psi_{\bf Ij}}\right)=u_{\bf
Ij}\prod_{i:j^i=0}\nu^i(I_0^i)
$$
for all ${\bf I}$, ${\bf j}$ (taking into account that $u_{\bf
Ij}=0$ if the corresponding $\Psi_{\bf Ij}=0$).

 }
\end{lemma}

\noindent{\it Proof}\qquad Let us note that the set of functions
$\{\Psi_{\bf Ij}\}$  is the set of all pairwise tensor products of
functions from the corresponding sets $\{\Psi_{I^ij_i}\}$.

In Lemma \ref{fix_gf} it was proven that $\{\Psi_{I^ij_i}\}$ is a
linearly independent complete set in $D(X^i)$.

Since $D(X)=\otimes_{i=1}^{n} D(X^i)$, the same will hold for the
set $\{\Psi_{\bf Ij}\}$.

Therefore any generalized function is determined by its action on
$\{\Psi_{\bf Ij}\}$. This finishes the proof of the lemma.

\bigskip

If all $j^i$, $i=1,\dots,n$ are non--zero, the corresponding
function $\Psi_{\bf Ij}$ is an ultrametric wavelet. If some $j^i$
are zero, then $\Psi_{\bf Ij}$ does not belong to the basis of
wavelets.

\begin{lemma}\label{lemma_series_1}
{\sl Assume that for a vertex ${\bf I}_0\in {\cal T}$, ${\bf
I}_0=I_0^1\times\dots\times I_0^n$ its measure is positive:
$\nu({\bf I}_0)=\prod_{i=1}^{n}\nu^i(I_0^i)>0$.

Then the series \be\label{the_series_1} u=\sum_{\bf Ij}u_{\bf
Ij}\otimes_{i:j^i\ne
0}\left(\Psi_{I^ij^i}-{1\over\nu^i(I^i_0)}\Psi_{\bf
Ij}\left(\chi_{I^i_0}\right)\right) \ee is the generalized function
$u\in D'(X)$ which satisfies the condition \be\label{init_cond_1}
u\left(\ov{\Psi_{\bf Ij}}\right)=u_{\bf
Ij}\prod_{i:j^i=0}\nu^i(I_0^i)\ee for all ${\bf I}$, ${\bf j}$
(again, taking into account that $u_{\bf Ij}=0$ if the corresponding
$\Psi_{\bf Ij}=0$).

Moreover an arbitrary $u\in D'(X)$ satisfying (\ref{init_cond_1})
has the form (\ref{the_series_1}). }
\end{lemma}

\noindent{\it Proof}\qquad To check that the series
(\ref{the_series_1}) is indeed a generalized function in $D'(X)$, we
have to prove that application of this series to any characteristic
function $\chi_{{\bf J}_0}$ gives a convergent series. We have

$$
u\left(\chi_{{\bf J}_0}\right)=\sum_{\bf Ij}u_{\bf Ij}\prod_{i:j^i=
0}\nu^i(J_0^i) \otimes_{i:j^i\ne
0}\left(\Psi_{I^ij^i}\left(\chi_{J^i_0}\right)-{\nu^i(J_0^i)\over\nu^i(I^i_0)}\Psi_{I^ij^i}
\left(\chi_{I^i_0}\right)\right)
$$

Using the same arguments as in Lemma \ref{lemma_series}, we get that
in the above series we will have nonzero terms only for those ${\bf
I,j}$ which for all $i=1,\dots,n$ satisfy one of the conditions:

1) $j^i\ne 0$, $J^i_0<I^i\le{\rm sup}\,(J^i_0,I^i_0)$;

2) $j^i\ne 0$, $I^i_0<I^i\le{\rm sup}\,(J^i_0,I^i_0)$;

3) $j^i= 0$, $I^i=I_0^i$.

Since the number of ${\bf I,j}$ satisfying the above conditions is
finite, the series (\ref{the_series_1}) converges in $D'(X)$.

Applying (\ref{the_series_1}) to $\ov{\Psi_{\bf Ij}}$ we get
(\ref{init_cond_1}).

An application of Lemma \ref{fix_gf_1} proves that an arbitrary
$u\in D'(X)$ satisfying (\ref{init_cond_1}) has the form
(\ref{the_series_1}). This finishes the proof of the lemma.

\bigskip

\section{Multidimensional pseudodifferential operators}

In this section we introduce multidimensional ultrametric
pseudodifferential operators. Let us consider the integral operators
(linear combinations of one--dimensional operators) in $L^2(X,\nu)$:
$$
T_i f=\int T_i({\rm
sup}(x^i,y^i))(f(x^1,\dots,x^i,\dots,x^n)-f(x^1,\dots,y^i,\dots,x^n))d\nu^i(y^i).
$$

We call a multidimensional ultrametric pseudodifferential operator
the following polynomial combination
\be\label{Tk}T=\sum_{k=1}^{m}\sum_{i_1\dots i_k=1}^n
a^{(k)}_{i_1\dots i_k} T_{i_1}\dots T_{i_k} \ee where
$a^{(k)}_{i_1\dots i_k}$ are complex numbers.

In particular, we will consider operators of the form
$$
T^{(1)}=\sum_{i=1}^n a_i T_i,\qquad T^{(2)}=\sum_{ij=1}^n a_{ij}
T_iT_j.
$$

Since the operators $T_i$ are diagonal in the wavelet basis
$$
T_i\Psi_{{\bf I} {\bf j}}=\lambda_{I^i}\Psi_{{\bf I} {\bf j}},
$$
the multidimensional pseudodifferential operators will be diagonal
in the basis of wavelets and the corresponding eigenvalues for
$\Psi_{{\bf I} {\bf j}}$ depend only on ${\bf I}$:
\be\label{lambda2} T\Psi_{{\bf I} {\bf
j}}=\left[\sum_{k=1}^{m}\sum_{i_1\dots i_k=1}^n a^{(k)}_{i_1\dots
i_k}\lambda_{I^{i_1}}\dots\lambda_{I^{i_k}}\right]\Psi_{{\bf I} {\bf
j}}=\lambda_{\bf I}\Psi_{{\bf I} {\bf j}} \ee

\begin{definition}{\sl
We will say that a vertex ${\bf I}\in {\cal T}$ is characteristic
for the multidimensional pseudodifferential operator $T$, if the
eigenvalue $\lambda_{\bf I}$ of the operator $T$ is equal to zero.}
\end{definition}

This definition might look strange, but actually it is a direct
analogue of the standard definition of a characteristic for real
differential operators. The vertex ${\bf I}=I^1\times\dots\times
I^n$ is characteristic when the corresponding vector of eigenvalues
$\Lambda=(\lambda_{I^i})$, $i=1,\dots,n$ lies in the kernel of the
multilinear form
$$
A(\Lambda)=\sum_{k=1}^{m}\sum_{i_1\dots i_k=1}^n a^{(k)}_{i_1\dots
i_k}\lambda_{I^{i_1}}\dots\lambda_{I^{i_k}}
$$
which defines the operator $T$.

This is the direct analogue of the real case, see Appendix 1 for
discussion. Moreover, in the simplest case where the space $X$ is
the product $Q_p^n$ of $p$--adic fields, the corresponding wavelet
in the simplest case will have the form of product of a character of
$Q_p^n$ and product of characteristic functions of balls. This can
be compared with the formulae (\ref{phi1}), (\ref{phi2}) of the
Appendix 1.

\section{Cauchy problem in the multidimensional case}

In the present section we investigate Cauchy problems for
ultrametric pseudodifferential equations in many dimensions. We
consider the following Cauchy problem: let us find $u$, satisfying
$$
Tu=f,
$$
where $T$ is an ultrametric pseudodifferential operator with the
eigenvalues $\lambda_{\bf I}$ in the wavelet basis, $f\in D'_0(X)$
lies in Lizorkin space of generalized functions, $u\in D'(X)$ is a
generalized function, which satisfies the initial conditions
(\ref{init_cond_1}) for some vertex ${\bf I}_0\in {\cal T}$ with the
positive measure $\nu({\bf I}_0)$.

We have the following necessary conditions for solvability of the
above Cauchy problem: if $\lambda_{\bf I}=0$ for some ${\bf I}\in
{\cal T}$ then $f_{\bf Ij}=f\left(\ov{\Psi_{\bf Ij}}\right)=0$ for
all corresponding ultrametric wavelets $\Psi_{\bf Ij}$.

We have the following ultrametric analogue of the Kovalevskaya
theorem.

\begin{theorem}
{\sl Let the multidimensional ultrametric pseudodifferential
operator $T$ have the form
$$
T=\sum_{k=1}^{m}\sum_{i_1\dots i_k=1}^n a^{(k)}_{i_1\dots i_k}
T_{i_1}\dots T_{i_k},
$$
where $T_i$ are one--dimensional ultrametric pseudodifferential
operators.

Consider the Cauchy problem \be\label{Cauchy} Tu=f \ee where $f\in
D'_0(X)$, with $u\in D'(X)$ satisfying the set of initial conditions
\be\label{init_cond_2} u\left(\ov{\Psi_{\bf Ij}}\right)=u_{\bf
Ij}\prod_{i:j^i=0}\nu^i(I_0^i) \ee for those ${\bf I}$, ${\bf j}$
for which at least one of the $j^i$, $i=1,\dots,n$ will be equal to
zero.

Assume that if ${\bf I}$ is characteristic, i.e. the corresponding
eigenvalue $\lambda_{\bf I}$ of the operator $T$ in the basis of
multidimensional ultrametric wavelets is equal to zero:
$$
\lambda_{\bf I}=\sum_{k=1}^{m}\sum_{i_1\dots i_k=1}^n
a^{(k)}_{i_1\dots i_k}\lambda_{I^{i_1}}\dots\lambda_{I^{i_k}}= 0,
$$
then $f_{\bf Ij}=f\left( \ov{\Psi_{{\bf I}{\bf j}}}\right)=0$ for
all ${\bf j}$ (here $\Psi_{{\bf I}{\bf j}}$ are ultrametric
wavelets, i.e. $j^i$ can not be equal to zero).

Then the solution $u$ of the Cauchy problem (\ref{Cauchy}) does
exist, is unique, belongs to $D'(X)$ and is given by the series over
ultrametric wavelets \be\label{u}\sum_{\bf Ij}u_{\bf
Ij}\otimes_{i:j^i\ne
0}\left(\Psi_{I^ij^i}-{1\over\nu^i(I^i_0)}\Psi_{\bf
Ij}\left(\chi_{I^i_0}\right)\right)+
$$
$$+\sum_{\bf Ij}{1\over\lambda_{\bf I}}f_{\bf Ij}\otimes_{i:j^i\ne
0}\left(\Psi_{I^ij^i}-{1\over\nu^i(I^i_0)}\Psi_{\bf
Ij}\left(\chi_{I^i_0}\right)\right)+
$$
$$
+\sum_{\bf Ij}u_{\bf Ij}\otimes_{i:j^i\ne
0}\left(\Psi_{I^ij^i}-{1\over\nu^i(I^i_0)}\Psi_{\bf
Ij}\left(\chi_{I^i_0}\right)\right). \ee Here summation in the first
sum runs over the initial conditions (the corresponding $u_{\bf Ij}$
are given by (\ref{init_cond_2})); summation in the second sum runs
over non--characteristic ${\bf I}$; summation in the third sum runs
over the characteristic ${\bf I}$ (the corresponding $u_{\bf Ij}$
are arbitrary).

}
\end{theorem}

\noindent{\it Proof}\qquad By Lemma \ref{lemma_series_1}, any
generalized function satisfying (\ref{init_cond_2}) has the form
(\ref{the_series_1}). Substituting this series into (\ref{Cauchy})
we get for the case where ${\bf I}$ is not characteristic
$$
u_{\bf Ij}={1\over\lambda_{\bf I}}f_{\bf Ij}.
$$
This implies for $u$ the expression \be\label{solution1} u=\sum_{\bf
Ij}u_{\bf Ij}\otimes_{i:j^i\ne
0}\left(\Psi_{I^ij^i}-{1\over\nu^i(I^i_0)}\Psi_{\bf
Ij}\left(\chi_{I^i_0}\right)\right)+
$$
$$+\sum_{\bf Ij}{1\over\lambda_{\bf I}}f_{\bf Ij}\otimes_{i:j^i\ne
0}\left(\Psi_{I^ij^i}-{1\over\nu^i(I^i_0)}\Psi_{\bf
Ij}\left(\chi_{I^i_0}\right)\right)+
$$
$$
+\sum_{\bf Ij}u_{\bf Ij}\otimes_{i:j^i\ne
0}\left(\Psi_{I^ij^i}-{1\over\nu^i(I^i_0)}\Psi_{\bf
Ij}\left(\chi_{I^i_0}\right)\right)\ee

Here the summation in the first sum runs over the initial
conditions, i.e. over those ${\bf I,j}$ for which at least one
$j^i$, $i=1,\dots,n$ is equal to zero. The corresponding $u_{\bf
Ij}$ are given by the initial conditions (\ref{init_cond_2}).

The summation in the second sum runs over the non--characteristic
vertices ${\bf I}$. The summation in the third sum runs over
characteristic vertices ${\bf I}$, and the corresponding $u_{\bf
Ij}$ are arbitrary.

This finishes the proof of the theorem.

\bigskip

We see that the possible source of problems for existence and
uniqueness of the solution for Cauchy problem is the case where some
vertex ${\bf I}$ is characteristic for the operator $T$.

\bigskip

\noindent{\bf Example}\qquad We shall discuss the case of the
ultrametric analogue of the propagating wave solution for the wave
equation. Consider the two dimensional case where $X=X^1\times X^2$,
$\nu=\nu^1\times\nu^2$, and the pseudodifferential operator has the
form
$$
T=T_1-T_2,
$$
which implies $\lambda_{\bf I}=\lambda_{I^1}-\lambda_{I^2}$.

Consider the Cauchy problem
$$
Tu=0,\qquad u\left(\chi_{{\bf I}_0}\right)=0.
$$

Assume that the vertex ${\bf I}$ is characteristic (i.e.
$\lambda_{I^1}=\lambda_{I^2}$) and ${\bf I}_0\bigcap {\bf
I}=\emptyset$. Then any $u\in V^0({\bf I})$ will be a solution of
the Cauchy problem.

In the case under consideration the solution of the Cauchy problem
does exist but it is not unique. We can consider this solution as an
ultrametric analogue of the propagating wave solution for the wave
equation. Note that this propagating wave propagates inside ${\bf
I}=I^1\times I^2$ and therefore is localized both in space and in
scale.

\section{Appendix 1: Characteristics}

We recall here the standard definitions. Consider the linear partial
differential equation of the order $m$: \be\label{diffequ}
\sum_{|\alpha|\le m}a_{\alpha}(x)\pa_{x}^{\alpha}u(x)=f(x) \ee Here
$\alpha=(\alpha_1,\dots,\alpha_n)$ is a multiindex,
$|\alpha|=\sum_{i=1}^n\alpha_i$ is the order of the multiindex,
$$
\pa_{x}^{\alpha}=\prod_{i=1}^n {\pa^{\alpha_i}\over\pa
x_i^{\alpha_i}}.
$$

We say that non--zero vector $\xi=(\xi_1,\dots,\xi_n)$ has a
characteristic direction in the point $x$, if the corresponding
characteristic polynomial of equation (\ref{diffequ}) is equal to
zero
$$
\sum_{|\alpha|= m}a_{\alpha}(x)\xi^{\alpha}=0,\qquad
\xi^{\alpha}=\prod_{i=1}^{n} \xi_i^{\alpha_i}.
$$

A hypersurface is said to be characteristic, if the normal vector to
this surface at any point has a characteristic direction.

Let us note that application of the terms of higher order of the
differential operator at the LHS of the equation (\ref{diffequ}) to
the function \be\label{phi1} \phi(x)=e^{i\sum_{l=1}^{n}\xi_lx_l} \ee
gives \be\label{phi2}D\phi(x)= i^{|\alpha|}\phi(x)\sum_{|\alpha|=
m}a_{\alpha}(x)\xi^{\alpha},\qquad D=\sum_{|\alpha|\le
m}a_{\alpha}(x)\pa_{x}^{\alpha}, \ee which is proportional to the
characteristic polynomial.

Note that the function $\phi$ is a character of the linear space
$R^n$ and is an eigenfunction (in the sense of generalized
functions) of the differentiation operators $\pa_{x_i}$, and the
characteristic polynomial in (\ref{phi2}) is a polynomial over the
corresponding eigenvalues.

\section{Appendix 2: Ultrametric analysis}

In this Section we summarize the results on ultrametric analysis,
which may  mainly be found in  \cite{Izv}, \cite{ACHA},
\cite{MathSbornik}. We discuss ultrametric wavelet analysis,
analysis of ultrametric pseudodifferential operators (PDO), and
ultrametric distribution theory.

\subsection{Ultrametric pseudodifferential operators}

\begin{definition}{\sl
An ultrametric space is a metric space with the ultrametric $d(x,y)$
(where $d(x,y)$ is called the distance between $x$ and $y$), i.e. a
function of two variables, satisfying the properties of positivity
and non degeneracy
$$
d(x,y)\ge 0,\qquad d(x,y)=0\quad \Longrightarrow\quad x=y;
$$
symmetricity
$$
d(x,y)=d(y,x);
$$
and the strong triangle inequality
$$
d(x,y)\le{\rm max }(d(x,z),d(y,z)),\qquad \forall x,y,z.
$$
}
\end{definition}

We say that an ultrametric space $X$ is regular, if this space
satisfies the following properties:

\medskip

1) The set of all the balls of nonzero diameter in $X$ is finite or
countable;

\medskip

2) For any decreasing sequence of balls $\{D^{(k)}\}$,
$D^{(k)}\supset D^{(k+1)}$, the diameters of the balls tend to zero;

\medskip

3) Any ball of non--zero diameter is a finite union of maximal
subballs.

\bigskip

Ultrametric spaces are dual to directed trees. Below we describe
some part of the duality construction.

For a regular ultrametric space $X$ consider the set ${\cal T}(X)$,
which contains all the balls in $X$ of nonzero diameters, and the
balls of zero diameter which are maximal subbals in balls of nonzero
diameters. This set possesses a natural structure of a directed
tree. Two vertices $I$ and $J$ in ${\cal T}(X)$ are connected by an
edge if the corresponding balls are ordered by inclusion, say
$I\supset J$ (i.e. one of the balls contain the other), and there
are no intermediate balls between $I$ and $J$.

The partial order in ${\cal T}(X)$ is defined by inclusion of balls,
this partial order is a direction. We recall that a partially
ordered set is a directed set (and a partial order is a direction),
if for any pair of elements there exists the unique supremum with
respect to the partial order.

A vertex $I$ of a directed tree has by definition a branching index
$p_I$, if the corresponding ball contains $p_I$ maximal subballs.

The supremum
$$
{\rm sup}(x,y)=I
$$
of the points $x,y\in X$ is the minimal ball $I$ in $X$, containing
both points.

Consider a $\sigma$--additive Borel measure $\nu$ with countable or
finite basis on a regular ultrametric space $X$. We study the
ultrametric pseudodifferential operator (shortly written PDO) of the
form considered in \cite{Izv}, \cite{ACHA}, \cite{MathSbornik},
i.e.:
$$
Tf(x)=\int T{({\rm sup}(x,y))}(f(x)-f(y))d\nu(y)
$$
Here $T{(I)}$ is some complex valued function on the tree ${\cal T}(
X)$.

\subsection{Ultrametric wavelets}

Build a basis in the space $L^2(X,\nu)$ of complex valued functions
on a regular ultrametric space $X$ which are quadratically
integrable with respect to the measure $\nu$. We will call this
basis the basis of ultrametric wavelets.

Denote by $V(I)$ the space of functions on $X$, generated by
characteristic functions of the maximal subballs in the ball $I$ of
nonzero diameter. Correspondingly, $V^0(I)$ is the subspace of
codimension 1 in $V(I)$ of functions with zero mean with respect to
the measure $\nu$. The spaces $V^0(I)$ for different $I$ are
orthogonal. The dimension of the space $V^0(I)$ is equal to $p_I-1$
(if $\nu(I_j)\ne 0$ for all maximal subballs $I_j$ in $I$).

We introduce in the space $V^0(I)$ some orthonormal basis
$\{\psi_{Ij}\}$. If the measures of all maximal subballs in $I$ are
positive, the index $j$ can take values $1,\dots,p_I-1$. The next
theorem shows how to construct the orthonormal basis in
$L^2(X,\nu)$, taking the union of bases $\{\psi_{Ij}\}$ in the
spaces $V^0(I)$ over all non minimal $I$ (equivalently, over all
balls $I$ of non--zero diameters).

\begin{theorem}\label{basisX}
{\sl 1) Let the measure $\nu(X)$ of the regular ultrametric space
$X$ is infinite. Then the set of functions $\{\psi_{Ij}\}$, where
$I$ runs over all non minimal vertices of the tree ${\cal T}(X)$ is
an orthonormal basis in $L^2(X,\nu)$.

2) Let the measure $\nu(X)$ of the regular ultrametric space $X$ is
finite and is equal to $\nu(X)=A$. Then the set of functions
$\{\psi_{Ij}, A^{-{1\over 2}}\}$, where $I$ runs over all non
minimal vertices of the tree ${\cal T}(X)$ is an orthonormal basis
in $L^2(X,\nu)$.
 }
\end{theorem}

The basis introduced in the present theorem will be called the basis
of ultrametric wavelets.

The next theorem shows that the basis of ultrametric wavelets is the
basis of eigenvectors for ultrametric pseudodifferential operators.

\begin{theorem}\label{04}{\sl Let the following series converge absolutely:
\be\label{seriesconverge} \sum_{J>R} T{(J)} (\nu(J)-\nu(J(R)))
<\infty, \ee for some ball $R$.

Then the ultrametric pseudodifferential operator
$$
Tf(x)=\int T{({\rm sup}(x,y))}(f(x)-f(y))d\nu(y)
$$
has a dense domain in $L^2(X,\nu)$ and ultrametric wavelets from
Theorem \ref{basisX} are eigenfunctions of $T$: \be\label{lemma2.1}
T\psi_{Ij}(x)=\lambda_I \psi_{Ij}(x) \ee with the eigenvalues:
\be\label{lemma4} \lambda_{I}=T{(I)} \nu(I)+\sum_{J>I} T{(J)}
(\nu(J)-\nu(J(I))) \ee Here $J(I)$ is the maximal subball in $J$
which contains $I$.

Also the operator $T$ maps constants into zero. }
\end{theorem}

\subsection{Distributions}

Here we discuss the spaces of (complex valued) test and generalized
functions (or distributions) on a regular ultrametric space $X$.
This construction is an analogue of the construction of the
Bruhat--Schwartz space in the $p$--adic case, which can be found in
\cite{VVZ}.

\begin{definition}\label{local_constant}{\sl A
function $f$ on an (ultrametric) space $X$ is called locally
constant, if for any arbitrary point $x\in X$ there exists a
positive number $r$ (depending on $x$), such that the function $f$
is constant on the ball with the center in $x$ and the radius $r$:
$$
f(x)=f(y),\qquad \forall y: d(x,y)\le r.
$$
}
\end{definition}

In particular, the characteristic function $\chi_I$ of a ball $I$ is
locally constant. Any locally constant function is continuous. The
next definition is an analogue of a known definition for the
Bruhat--Schwartz space in the $p$--adic case.

\begin{definition}\label{test_functions}{\sl
The space of test functions $D(X)$ on an ultrametric space $X$ is
defined as the space of locally constant functions with compact
support.}
\end{definition}

We remind that a filtration of a set $A$ by a partially ordered set
$B$ is a map $\phi$ which maps $B$ to the partially ordered (by
inclusion) set of subsets in $A$, and $\phi$ satisfies the following
conditions:

1) the map $\phi$ conserves the partial order, i.e. if $b<b'$,
$b,b'\in B$, then there exists an embedding of $\phi(b)$ into
$\phi(b')$;

2) any element of the set $A$ lies in the image of some element of
$B$.

In the case of filtrated linear spaces we will consider filtrations
by families of linear subspaces, i.e. an embedding  of $\phi(b)$
into $\phi(b')$ for $b<b'$, $b,b'\in B$ should be linear.

Consider the tree ${\cal T}={\cal T}(X)$ of balls in an ultrametric
space $X$. Introduce the filtration of the space $D(X)$ of test
functions by finite dimensional linear subspaces  $D({\cal S})$,
where ${\cal S}\subset {\cal T}$ is a finite directed subtree of the
following ''regular'' form:

\begin{definition}\label{wave_type}
{\sl A finite subtree ${\cal S}$ in a directed tree ${\cal T}(X)$ is
called of the regular type, iff:

1) If ${\cal S}$ contains the balls $I$ and $J$, then it contains
${{\rm sup}(I,J)}$;

2) If ${\cal S}$ contains the balls $I$ and $J$: $I\subset J$, then
${\cal S}$ contains all the balls $L$: $I\subset L\subset J$;

3) If ${\cal S}$ contains the balls $I$ and $J$, where $J$ is a
maximal subball in $I$, then it contains all the maximal subballs in
$I$.
 }
\end{definition}

We denote by $\chi_J$ a characteristic function of a ball $J$.

\begin{definition}\label{D(S)}
{\sl For a subtree ${\cal S}\subset {\cal T}$ of the regular type
consider the space $D({\cal S})$, which is the linear span of
characteristic functions $\chi_J$ with $J\in {\cal S}$. }
\end{definition}

The space $D({\cal S})$ has the natural topology (which can be
described as the topology of pointwise convergence), since $D({\cal
S})$ is finite dimensional.  The space $D(X)$ of test functions on
$X$ is the inductive limit of spaces $D({\cal S})$:
\be\label{inductive} D(X)=\lim\,{\rm ind}_{{\cal S}\to {\cal T}}\,
D({\cal S})\ee

Since all spaces $D({\cal S})$ are finite dimensional, the
restriction of the topology of $D({\cal S})$ to any subspace
$D({\cal S}_0)$, ${\cal S}\supset {\cal S}_0$, coincides with the
original topology of $D({\cal S}_0)$. Thus the inductive limit
(\ref{inductive}) is, in fact, the {\it strict inductive limit}
\cite{Schaefer} p.57. By proposition 6.5, \cite{Schaefer}, p.59, a
set $B$ in a strict inductive limit of a countable family of locally
convex spaces $\{E_n\}$ is bounded iff there exists $n$ such that
$B\subset E_{n}$ and bounded in it. This implies the proposition
below.

\begin{proposition}\label{topology}{\sl The sequence
$\{f_n\}\in D(X)$ converges, if this sequence lies in some subspace
$D({\cal S})$ of the filtration and converges (and the convergence
in the finite dimensional space $D({\cal S})$ is defined uniquely).}
\end{proposition}

By corollary of theorem 7.4, \cite{Schaefer}, p.103, we have the
following proposition.

\begin{proposition}
{\sl The space $D(X)$ is nuclear.}
\end{proposition}

\begin{definition}\label{generalized_functions}{\sl
A distribution (or a generalized function) on $X$ is a linear
functional on the space $D(X)$ of test functions.}
\end{definition}

It is easy to see that this functional automatically will be
continuous (since convergence in the space of test functions is
defined through the convergence in finite dimensional subspaces).
The linear space of generalized functions will be denoted by
$D'(X)$. The convergence in $D'(X)$ is defined as a weak convergence
of functionals. Thus $D'(X)$ is conjugated to $D(X)$ with the weak
topology.

\bigskip\bigskip

\noindent{\bf Acknowledgments}\qquad One of the authors (S.K.) would
like to thank I.V.Volovich, V.S.Vla\-di\-mi\-rov, A.Yu.Khrennikov
and V.M.Shelkovich for fruitful discussions and valuable comments.
He gratefully acknowledges being partially supported by the grant
DFG Project 436 RUS 113/809/0-1, by the grants of The Russian
Foundation for Basic Research  RFFI 05-01-04002-NNIO-a and RFFI
05-01-00884-a, by the grant of the President of Russian Federation
for the support of scientific schools NSh 6705.2006.1 and by the
Program of the Department of Mathematics of Russian Academy of
Science ''Modern problems of theoretical mathematics''.

S.K. is grateful to IZKS (The Interdisciplinary Center for Complex
Systems) of the University of Bonn for kind hospitality.

\end{document}